\documentclass[12pt]{article}
\usepackage{times}
\usepackage{epsfig}
\usepackage{axodraw}
\usepackage{amsmath}
\usepackage{amsfonts}

\def\beq{\begin{equation}}
\def\eeq{\end{equation}}
\def\bea{\begin{eqnarray}}
\def\eea{\end{eqnarray}}

\def\nn{\nonumber}
\def\Eq#1{Eq.~(\ref{#1})}

\def\td#1{\widetilde{\delta}\left(#1\right)}

\begin{document}

\begin{titlepage}
\renewcommand{\thefootnote}{\fnsymbol{footnote}}
\begin{flushright}
LPN12-057 \\
IFIC/12-44 \\
DESY 12-208
\end{flushright}
\par \vspace{10mm}
\begin{center}
{\LARGE \bf
Tree--Loop Duality Relation beyond single poles}
\end{center}
\par \vspace{2mm}
\begin{center}
{\bf 
Isabella Bierenbaum~$^{(a)}$\footnote{E-mail: isabella.bierenbaum@desy.de}, 
Sebastian Buchta~$^{(b)}$\footnote{E-mail: sbuchta@ific.uv.es}, 
Petros Draggiotis~$^{(b)}$\footnote{E-mail: petros.drangiotis@ific.uv.es},
Ioannis Malamos~$^{(b)}$\footnote{E-mail: ioannis.malamos@ific.uv.es}and
Germ\'an Rodrigo~$^{(b)}$\footnote{E-mail: german.rodrigo@ific.uv.es}
}
\vspace{5mm}

${}^{(a)}$ II. Institut f\"{u}r Theoretische Physik, Universit\"{a}t Hamburg,\\
Luruper Chaussee 149, 22761, Hamburg, Germany
\\
\vspace*{2mm}
${}^{(b)}$ Instituto de F\'{\i}sica Corpuscular, 
Universitat de Val\`{e}ncia -- 
Consejo Superior de Investigaciones Cient\'{\i}ficas, 
Parc Cient\'{\i}fic, E-46980 Paterna (Valencia), Spain. \\
\vspace*{2mm}

\end{center}

\par \vspace{2mm}
\begin{center} {\large \bf Abstract} \end{center}
\begin{quote}
We develop the Tree--Loop Duality Relation for two-- and three--loop integrals with multiple
identical propagators (multiple poles). This is the extension of the Duality 
Relation for single poles and multi--loop integrals derived in previous publications. We prove
a generalization of the formula for single poles to multiple poles and we develop a strategy for dealing
with higher-order pole integrals by reducing them to single pole integrals using Integration
By Parts.
\end{quote}

\par \vspace{5mm}

\vspace*{\fill}
\end{titlepage}

\setcounter{footnote}{0}
\renewcommand{\thefootnote}{\fnsymbol{footnote}}

\section{Introduction}

The last 15 years have seen extraordinary progress in the analytical and numerical computation of cross sections
in the Standard Model at one-- and two--loops and even higher orders. Once considered the bottleneck of numerical applications, higher-
than-tree-level corrections exhibit high complexity, partly  due to the enormous number of Feynman diagrams needed, sometimes
numbered in hundreds or even thousands, for important cross sections. Important efforts have been devoted to developing 
efficient methods able to boost forward the calculational capability both at the multi--leg and
the multi--loop frontier.  Today, $2\to 4$ processes at next--to--leading order (NLO), either from
Unitarity based methods,~\cite{Berger:2009zg,Melnikov:2009wh,Bevilacqua:2009zn}, or from a
more traditional Feynman diagrammatic approach,~\cite{Bredenstein:2010rs}, are affordable and are even becoming 
standardized.  There has also been a lot of progress concerning next--to--next--to leading order (NNLO) calculations
\cite{Bolzoni:2010xr,Catani:2010en,Catani:2009sm,Anastasiou:2007mz}.

The importance of higher order corrections cannot be overstated. While leading--order (LO) predictions 
of multi--particle processes at hadron colliders in perturbative Quantum Chromodynamics (pQCD) provide, in general, a
rather poor description of experimental data, NLO is the first order at which normalizations, 
and in some cases, the shapes, of cross sections can be considered reliable ~\cite{Binoth:2010ra}. 
NNLO, besides improving the determination of normalizations and shapes, 
is also generally accepted to provide the first serious estimate of the theoretical uncertainty in pQCD.
Despite the relatively smaller coupling, electroweak (EW) radiative NLO corrections might
also be sizable at the LHC,~\cite{Denner:2009gj,Kuhn:2007cv}.
Precision theoretical predictions for background and signal multi--particle hard scattering processes, 
in the SM and beyond, are mandatory for the phenomenological interpretation of experimental data, and thus to achieve a
successful exploitation of the LHC physics programme.

In \cite{Catani:2008xa}, a novel method was developed for the calculation of multileg one--loop
amplitudes. Called the Duality Theorem, it applies directly the Cauchy Residue Theorem to one--loop integrals. The result
can be represented by single cuts to Feynman diagrams, integrated over a modified phase space. 
(Note also \cite{CaronHuot:2010zt} where the author uses retarded boundary conditions to obtain
a duality between loop and tree graphs).   
The Duality Theorem was
extended in \cite{Bierenbaum:2010cy} beyond the one--loop level, to two-- and three--loops and it was shown
how it can be extended to an arbitrary number of loops. The main feature and advantage of this approach is that 
at any number of loops, an amplitude can be written as a sum of tree-level objects, obtained after making all possible
cuts to the lines of a Feynman diagram, one cut per loop and integrated over a measure that closely resembles 
the phase space of the corresponding real corrections. This modified phase-space, raises the intriguing possibility
that virtual and real corrections can be brought together under a common integral and treated with Monte-Carlo techniques
at the same time. In these papers the Duality Theorem was developed for diagrams and their integrals, that 
do not possess identical propagators. This possibility does not appear at one--loop for a convenient 
choice of gauge \cite{Catani:2008xa}, but it is always present
for higher order corrections. Identical propagators possess higher than single poles and the Duality Theorem 
developed so far, which
is based on the Cauchy formula for single poles, must be extended to accomodate for this new feature. 

The purpose of this work is to extend the Duality Theorem to graphs and integrals with multiple poles at two-- and 
three--loops and to present a procedure, a strategy for dealing with higher poles in an amplitude calculation that
retains the features and advantages of the Duality theorem as detailed in \cite{Bierenbaum:2010cy}.
The paper is organized as follows: In Section 2 we recall the basic definitions and results concerning
the Duality Theorem at one-- and two--loops and beyond, for integrals with single poles. In
Section 3 we derive in detail the extension of the Duality Theorem to integrals with double and multiple propagators, which
exhibit multiple poles in the complex plane. In section 4 we present an alternative method to deal with multiple
poles at two-- and three--loops which retains the basic advantages of the Duality Theorem. It is based on Integration 
By Parts relations that allow to rewrite integrals with multiple poles in terms of integrals involving only single poles.

\section{Duality relation at one-- and two--loops}
\label{sec:one-loop}
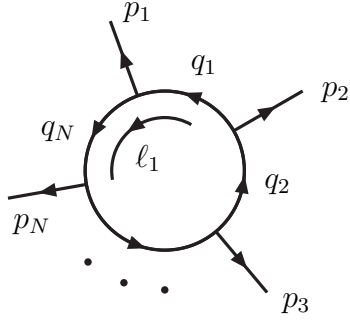
\begin{figure}[t]
\begin{center}
\vspace*{8mm}
\begin{picture}(120,110)(0,-10)
\SetWidth{1.2}
\BCirc(50,50){30}
\ArrowArc(50,50)(30,110,190)
\ArrowArc(50,50)(30,190,-50)
\ArrowArc(50,50)(30,-50,30)
\ArrowArc(50,50)(30,30,110)
\ArrowArc(50,50)(20,60,190)
\ArrowLine(39.74,78.19)(29.48,106.38)
\ArrowLine(75.98,65)(101.96,80)
\ArrowLine(69.28,27.01)(88.56,4.03)
\ArrowLine(20.45,44.79)(-9.09,39.58)
\Vertex(21.07,15.53){1.4}
\Vertex(34.60,7.71){1.4}
\Vertex(50,5){1.4}
\Text(44,55)[]{$\ell_1$}
\Text(40,110)[]{$p_1$}
\Text(65,90)[]{$q_1$}
\Text(115,80)[]{$p_2$}
\Text(93,45)[]{$q_2$}
\Text(10,65)[]{$q_N$}
\Text(0,30)[]{$p_N$}
\Text(100,0)[]{$p_3$}
\end{picture}
\end{center}
\vspace*{-6mm}
\caption{\label{f1loop} 
{\em Momentum configuration of the one--loop $N$--point scalar integral.}}
\end{figure}

A general one--loop $N$--leg, scalar integral, such as the one shown in Fig.~\ref{f1loop}, is written as:
\beq
\label{Ln}
L^{(1)}(p_1, p_2, \dots, p_N) =
\int_{\ell_1} \, \prod_{i=1}^{N} \, G_F(q_i)~,
\eeq
where 
\beq
G_F(q_i)=\frac{1}{q_i^2-m_i^2+i0}
\eeq
is the Feynman propagator.
The four--momenta of the external legs are denoted $p_i$, $i =
\{1,2,\ldots N\}$. All are taken as outgoing and ordered clockwise.  The
loop momentum is $\ell_1$, which flows anti--clockwise. The momenta of the
internal lines $q_i$, are defined as
\beq
\label{defqi}
q_i = \ell_1 + p_{1,i}~, \qquad i \in \alpha_1 = \{1,2,\ldots N\}~, \qquad p_{1,i}=p_1+\ldots +p_i~.
\eeq 
Throughout this paper we use the short-hand notation
\beq
\label{tdp}
\td{q_i} \equiv 2 \pi \, i \, \theta(q_{i,0}) \, \delta(q_i^2-m_i^2) 
= 2 \pi \, i \, \delta_+(q_i^2-m_i^2)~,
\quad \int_{\ell_i} \bullet =-i \int \frac{d^d \ell_i}{(2\pi)^{d}} \; \bullet
\eeq
where the subscript $+$ of $\delta_+$ refers to the on--shell mode with
positive definite energy, $q_{i,0}\geq 0$. Hence, the phase--space integral of
a physical particle with momentum $q_i$, i.e., an on--shell particle with
positive--definite energy, $q_i^2=m_i^2$, $q_{i,0}\geq 0$, reads:
\beq
\label{psm}
\int \frac{d^d q_i}{(2\pi)^{d-1}} \, \theta(q_{i,0}) \, \delta(q_i^2-m_i^2) 
\; \cdots \equiv \int_{q_i} \td{q_i} \; \cdots~.
\eeq
It was shown in \cite{Catani:2008xa,Bierenbaum:2010cy} that using the Cauchy residue theorem
the one--loop integral can be written in the form:
\bea
\label{oneloopduality}
L^{(1)}(p_1, p_2, \dots, p_N) 
&=& - \sum_{i} \, \int_{\ell_1} \; \td{q_i} \,
\prod_{\substack{j=1 \\ j\neq i}}^{N} \,G_D(q_i;q_j)~,
\eea 
where
\beq
G_D(q_i;q_j) = \frac{1}{q_j^2 -m_j^2 - i0 \,\eta (q_j-q_i)}~,
\eeq
is the so--called dual propagator, as defined in Ref.~\cite{Catani:2008xa},
with $\eta$ a {\em future--like} vector,
\beq
\label{etadef}
 \eta_0 \geq 0, 
\; \eta^2 = \eta_\mu \eta^\mu \geq 0 \;\;,
\eeq
i.e.,~a $d$--dimensional vector that can be either light--like $(\eta^2=0)$ or
time--like $(\eta^2 > 0)$ with positive definite energy $\eta_0$. 
The result in \Eq{oneloopduality}, contrary to the Feynman--Tree theorem (FTT) \cite{Feynman:1963ax,F2},  contains
only single--cut integrals. Multiple--cut integrals, like those that appear in
the FTT, are absent thanks to modifying the original $+i0$ prescription of the
uncut Feynman propagators by the new prescription $- i0 \,\eta
(q_j-q_i)$, which is named the `dual' $i0$ prescription or, briefly, the
$\eta$ prescription.
The dual $i0$ prescription arises from the fact that the original Feynman
propagator $G_F(q_j)$ is evaluated at the {\em complex} value of the loop
momentum $\ell_1$, which is determined by the location of the pole at
$q_i^2-m_i^2+i0 = 0$.  The $i0$ dependence of the pole of $G_F(q_i)$ modifies
the $i0$ dependence in the Feynman propagator $G_F(q_j)$, leading to the total
dependence as given by the dual $i0$ prescription. The presence of the vector
$\eta_\mu$ is a consequence of using the residue theorem and the fact that the
residues at each of the poles are not Lorentz--invariant quantities.  The
Lorentz--invariance of the loop integral is recovered after summing over all
the residues. Furthermore, in the one--loop case, the momentum difference
$\eta(q_j-q_i)$ is independent of the integration momentum $\ell_1$, and only
depends on the momenta of the external legs (cf. Eq.~(\ref{defqi})).


The extension of the Duality Theorem to two loops has been discussed in detail in 
\cite{Bierenbaum:2010cy,isabella}. Here we recall the basic points.
We extend the definition of propagators 
of single momenta to combinations of propagators of
sets of internal momenta. Let $\alpha_k$ be any set of internal momenta
$q_i,q_j$ with $i,j \in \alpha_k$.  We then define Feynman and dual
propagator functions of this set $\alpha_k$ in the following way:
\beq
\label{eq:multi}
G_{F} ( \alpha_k) = \prod_{i \in \alpha_k} G_{F}( q_i)~, \qquad
G_D( \alpha_k) = \sum_{i \in \alpha_k} \, \td{ q_{i}} \, 
\prod_{\substack{j \in \alpha_k \\ j \neq i }} \, G_D( q_i; q_j)~.
\eeq
By definition, $G_D(\alpha_k)=\td{q_i}$, when $\alpha_k = \{i\}$ and thus
consists of a single four momentum. At one--loop order, $\alpha_k$ is
naturally given by all internal momenta of the diagram which depend on the
single integration loop momentum $\ell_1$, $\alpha_k=\{1,2,\ldots, N\}$.
However, let us stress that $\alpha_k$ can in principle be any set of
internal momenta. At higher order loops, e.g., several integration loop momenta
are needed, and we can define \emph{several loop lines}  $\alpha_k$ to label all the
internal momenta (cf. \Eq{lines}) where \Eq{eq:multi} will be used for these
loop lines or unifications of these. We also define:
\beq
\label{eq:multiminus}
G_D(-\alpha_k) = \sum_{i \in \alpha_k} \, \td{-q_{i}} \, 
\prod_{\substack{j \in \alpha_k \\ j \neq i }} \, G_D(-q_i;-q_j)~, 
\eeq
where the sign in front of $\alpha_k$ indicates that we have reversed the
momentum flow of all the internal lines in $\alpha_k$. For Feynman
propagators, moreover, $G_F(-\alpha_k)=G_F(\alpha_k)$.
Using this notation the following relation holds for any set of
internal momenta $\alpha_k$:
\beq
G_A(\alpha_k) = G_F(\alpha_k) + G_D(\alpha_k)~,
\label{eq:relevant}
\eeq
where $G_A(q_i)$ is the advanced propagator:
\beq
G_A(q_i) = \frac{1}{q_i^2-m_i^2-i0\,q_{i,0}}~,
\eeq
and 
\beq
G_{A} ( \alpha_k) = \prod_{i \in \alpha_k} G_{A}( q_i)~.
\eeq
The proof of \Eq{eq:relevant} can be found in Ref. \cite{Bierenbaum:2010cy}.
Note that individual
terms in $G_D(\alpha_k)$ depend on the dual vector $\eta$, but the sum over
all terms contributing to $G_D(\alpha_k)$ is independent of it.  
Another crucial relation for the following is given by a formula that
allows to express the dual function of a set of momenta in terms of chosen
subsets. Considering the following set $\beta_N \equiv \alpha_1 \cup ... \cup
\alpha_N$, where $\beta_N$ is the unification of various subsets
$\alpha_i$, we can obtain the relation: 
\bea
\label{eq:GAinGDGeneralN}
G_D(\alpha_1 \cup \alpha_2 \cup ... \cup \alpha_N) 
&=& \sum_{\substack{\beta_N^{(1)} \cup     \beta_N^{(2)}  = \beta_N}} \,
\prod_{\substack{i_1\in \beta_N^{(1)}}} \, G_D(\alpha_{i_1}) \,
\prod_{\substack{i_2\in \beta_N^{(2)}}} \,G_F(\alpha_{i_2})\,.  
\eea 
The sum runs over all partitions of $\beta_N$ into exactly two blocks
$\beta_N^{(1)}$ and $\beta_N^{(2)}$ with elements $\alpha_i,\linebreak i\in
\{1,...,N\}$, where, contrary to the usual definition, we include the case:
$\beta_N^{(1)} \equiv \beta_N$, $\beta_N^{(2)} \equiv \emptyset$.
For the case of $N=2$, e.g.,
where $\beta_2 \equiv \alpha_1 \cup \alpha_2$, we have:
\beq
\label{eq:twoGD}
G_D(\alpha_1 \cup \alpha_2) 
= G_D(\alpha_1) \, G_D(\alpha_2)
+ G_D(\alpha_1) \, G_F(\alpha_2)
+ G_F(\alpha_1) \, G_D(\alpha_2)~.  
\eeq
Naturally it holds that:
\beq
G_F(\alpha_1 \cup \alpha_2 \cup ... \cup \alpha_N)= 
\prod_{i=1}^N G_F(\alpha_i)~.
\eeq
Since in general relation (\ref{eq:GAinGDGeneralN}) holds for any
set of basic elements $\alpha_i$ which are sets of internal momenta,
one can look at these expressions in different ways, depending on the given
sets and subsets considered. If we define, for example, the basic subsets
$\alpha_i$ to be given by single momenta $q_i$, and since in that case
$G_D(q_i) = \td{q_i}$, Eq.~(\ref{eq:GAinGDGeneralN}) then denotes a sum over
all possible differing m--tuple cuts for the momenta in the set $\beta_N$,
while the uncut propagators are Feynman propagators. These cuts start from
single cuts up to the maximal number of cuts given by the term where all the
propagators of the considered set are cut.
Using this notation, the Duality Theorem at one--loop can be written in
the compact form:
\beq
\label{eq:simpledual}
L^{(1)}(p_1, p_2, \dots, p_N) = - \int_{\ell_1} G_D(\alpha_1)~, 
\eeq
where $\alpha_1$ as in \Eq{defqi} labels {\it all} internal momenta $q_i$.
In this way, we directly obtain the duality relation between one--loop
integrals and single--cut phase--space integrals and hence \Eq{eq:simpledual}
can also be interpreted as the application of the Duality Theorem to the given set
of momenta $\alpha_1$. It obviously agrees, at one loop, with
\Eq{oneloopduality}. 

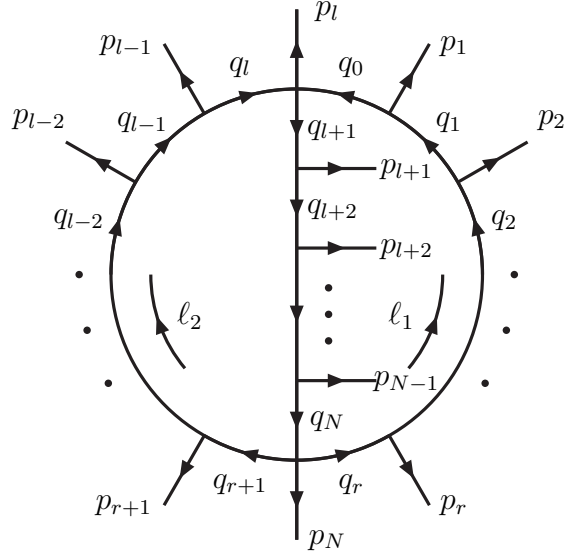
\begin{figure}[t]
\begin{center}

\begin{picture}(200,200)(0,-10)
\SetWidth{1.2}

\BCirc(100,100){70}
\Line(100,200)(100,0)
\ArrowLine(100,170)(100,200)
\ArrowLine(100,30)(100,0)

\ArrowArcn(100,100)(55,220,180)
\ArrowArc(100,100)(55,-40,0)

\ArrowLine(100,60)(130,60)
\ArrowLine(100,140)(130,140)
\ArrowLine(100,110)(130,110)
\Vertex(112,95){1.4}
\Vertex(112,85){1.4}
\Vertex(112,75){1.4}

\ArrowLine(135,161)(150,187)
\ArrowLine(161,135)(187,150)
\ArrowLine(135,39)(150,13)
\Vertex(179,79){1.4}
\Vertex(171,59){1.4}
\Vertex(182,100){1.4}

\ArrowLine(65,161)(50,187)
\ArrowLine(39,135)(13,150)
\ArrowLine(65,39)(50,13)
\Vertex(21,79){1.4}
\Vertex(29,59){1.4}
\Vertex(18,100){1.4}

\Text(160,187)[]{$p_1$}
\Text(197,159)[]{$p_2$}
\Text(160,13)[]{$p_r$}
\Text(112,0)[]{$p_N$}
\Text(35,13)[]{$p_{r+1}$}
\Text(112,200)[]{$p_{l}$}
\Text(142,140)[]{$p_{l+1}$}
\Text(142,110)[]{$p_{l+2}$}
\Text(142,60)[]{$p_{N-1}$}
\Text(36,187)[]{$p_{l-1}$}
\Text(3,159)[]{$p_{l-2}$}

\ArrowArc(100,100)(70,60,90)
\ArrowArc(100,100)(70,30,60)
\ArrowArc(100,100)(70,0,30)
\ArrowArc(100,100)(70,-90,-60)

\ArrowArcn(100,100)(70,120,90)
\ArrowArcn(100,100)(70,150,120)
\ArrowArcn(100,100)(70,180,150)
\ArrowArcn(100,100)(70,270,240)

\ArrowLine(100,170)(100,140)
\ArrowLine(100,140)(100,110)
\ArrowLine(100,110)(100,60)
\ArrowLine(100,60)(100,30)

\Text(121,179)[]{$q_0$}
\Text(158,158)[]{$q_1$}
\Text(179,121)[]{$q_2$}
\Text(121,21)[]{$q_r$}
\Text(79,179)[]{$q_l$}
\Text(42,158)[]{$q_{l-1}$}
\Text(18,121)[]{$q_{l-2}$}
\Text(79,21)[]{$q_{r+1}$}
\Text(114,155)[]{$q_{l+1}$}
\Text(114,125)[]{$q_{l+2}$}
\Text(112,45)[]{$q_N$}
\Text(140,85)[]{$\ell_1$}
\Text(60,85)[]{$\ell_2$}

\end{picture}
\end{center}
\caption{\label{f2loop}
{\em Momentum configuration of the two--loop $N$--point scalar integral.}}
\end{figure}

We now turn to the general two--loop master diagram, as presented in
Fig.~\ref{f2loop}.  Again, all external momenta $p_i$ are taken as outgoing,
and we have $p_{i,j}=p_i+p_{i+1}+\ldots +p_j$, with momentum conservation
$p_{1,N} = 0$. The label $i$ of the external momenta is defined modulo $N$,
i.e., $p_{N+i} \equiv p_{i}$. In the two--loop case, unlike at the one--loop order, the number of
external momenta might differ from the number of internal momenta. The loop momenta
are $\ell_1$ and $\ell_2$, which flow anti--clockwise and clockwise
respectively.  The momenta of the internal lines are denoted by $q_i$ and are
explicitly given by
\beq
\label{defqi2l}
q_i = \left\{
\begin{tabular}{ll}
$\ell_1+p_{1,i}$ & , $i \in \alpha_1$ \\
$\ell_2+p_{i,l-1}$ & , $i \in \alpha_2$ \\
$\ell_1+\ell_2+p_{i,l-1}$ &  , $i\in \alpha_3$ ~,
\end{tabular}
\right.  \eeq 
where $\alpha_k$, with $k=1,2,3$, are defined as the set of lines, propagators
respectively, related to the momenta $q_i$, for the following ranges of $i$:
\beq
\label{lines}
\alpha_1\equiv \{0,1,...,r\}~,\qquad \alpha_2\equiv \{r+1,r+2,...,l\}~,
\qquad \alpha_3\equiv \{l+1,l+2,...,N\}~.
\eeq
In the following, we will use $\alpha_k$
for denoting a set of indices or the set of the corresponding internal momenta
synonymously. Furthermore, we will refer to these lines often simply as the
``loop lines''.

We shall now extend the duality theorem to the two--loop case, 
by applying \Eq{eq:simpledual} iteratively. We consider first, in 
the most general form, a set of several loop lines $\alpha_1$ to $\alpha_N$
depending on the same integration momentum $\ell_i$, and find
\beq
\label{eq:ApplyDual}
\int_{\ell_i} \; G_F(\alpha_1 \cup \alpha_2 \cup ... \cup \alpha_N) 
= 
- \int_{\ell_i} \; G_D(\alpha_1 \cup \alpha_2 \cup ... \cup \alpha_N)~,
\eeq
which states the application of the duality theorem, \Eq{eq:simpledual}, to the set
of loop lines belonging to the same loop. \Eq{eq:ApplyDual} is the
generalization of the Duality Theorem found at one--loop to a single loop of a
multi--loop diagram. Each subsequent application of the Duality Theorem to
another loop of the same diagram will introduce an extra single cut, and by
applying the Duality Theorem as many times as the number of loops, a given
multi--loop diagram will be opened to a tree--level diagram. The Duality
Theorem, \Eq{eq:ApplyDual}, however, applies only to Feynman propagators, and
a subset of the loop lines whose propagators are transformed into dual
propagators by the application of the Duality Theorem to the first loop might
also be part of the next loop (cf., e.g., the ``middle'' line belonging to
$\alpha_3$ in Fig.~\ref{f2loop}). The dual function of the unification of
several subsets can be expressed in terms of dual and Feynman functions of the
individual subsets by using \Eq{eq:GAinGDGeneralN} (or \Eq{eq:twoGD}) though,
and we will use these expressions to transform part of the dual propagators
into Feynman propagators, in order to apply the Duality Theorem to the second
loop. Therefore, applying \Eq{eq:ApplyDual} to the loop with loop momentum $\ell_1$, 
reexpressing the result via \Eq{eq:twoGD} 
in terms of dual and Feynman propagators and applying \Eq{eq:ApplyDual} to the second
loop with momentum $\ell_2$, we obtain the duality relation at two loops in the form:
\bea
\label{AdvDual}
&& \!\!\!\!\!\!\!\!\!\!
L^{(2)}(p_1, p_2, \dots, p_N)  \nn \\
&& =   \int_{\ell_1} \int_{\ell_2} \, \left\{
- G_D(\alpha_1) \, G_F(\alpha_2) \, G_D(\alpha_3) 
+ G_D(\alpha_1) \, G_D(\alpha_2\cup \alpha_3)
+ G_D(\alpha_3) \, G_D(-\alpha_1\cup \alpha_2) \right\}~. \nn \\
\eea 
This is the dual representation of the two--loop scalar integral as a function
of double--cut integrals only, since all the terms of the integrand in
\Eq{AdvDual} contain exactly two dual functions as defined in \Eq{eq:multi}.
The integrand in \Eq{AdvDual} can then be reinterpreted as the sum over
tree--level diagrams integrated over a two--body phase--space.

The integrand in \Eq{AdvDual}, however, contains several dual functions of two
different loop lines, and hence dual propagators whose dual $i0$ prescription
might still depend on the integration momenta. This is the case for dual
propagators $G_D(q_i;q_j)$ where each of the momenta $q_i$ and $q_j$ belong to
different loop lines. If both momenta belong to the same loop line the
dependence on the integration momenta in $\eta(q_j-q_i)$ obviously cancels,
and the complex dual prescription is determined by external momenta only. The
dual prescription $\eta(q_j-q_i)$ can thus, in some cases, change sign within
the integration volume, therefore moving up or down the position of the poles
in the complex plane. To avoid this, we should reexpress the dual
representation of the two--loop scalar integral in \Eq{AdvDual} in terms of
dual functions of single loop lines. This transformation was unnecessary at
one--loop because at the lowest order all the internal momenta depend on the
same integration loop momenta; in other words, there is only a single loop
line.

Inserting \Eq{eq:twoGD} in \Eq{AdvDual} and reordering some terms, we arrive
at the following representation of the two--loop scalar integral
\bea
\label{AdvDualstar}
&& \!\!\!\!\!\!\!\!\!\!
L^{(2)}(p_1, p_2, \dots, p_N)  \nn \\
&& =   \int_{\ell_1} \int_{\ell_2} \, \left\{
  G_D(\alpha_1)  \, G_D(\alpha_2) \, G_F(\alpha_3) 
+ G_D(-\alpha_1) \, G_F(\alpha_2) \, G_D(\alpha_3)
+ G^*(\alpha_1)  \, G_D(\alpha_2) \, G_D(\alpha_3) \right\}~, \nn \\
\eea 
where
\beq
\label{Gstar1}
G^*(\alpha_1) \equiv G_F(\alpha_1) + G_D(\alpha_1) + G_D(-\alpha_1)~.
\eeq
In \Eq{AdvDualstar}, the
$i0$ prescription of all the dual propagators depends on external momenta
only. Through \Eq{Gstar1}, however, \Eq{AdvDualstar} contains also triple
cuts, given by the contributions with three $G_D(\alpha_k)$. The triple cuts
are such that they split the two--loop diagram into two disconnected
tree--level diagrams. By definition, however, the triple cuts
are such that there is no more than one cut per loop line $\alpha_k$. Since
there is only one loop line at one--loop, it is also clear why we did not
generate disconnected graphs at this loop order.
For a higher number of loops, we expect to
find at least the same number of cuts as the number of loops, and topology-dependent 
disconnected tree diagrams built by cutting up to all the loop lines
$\alpha_k$.
These results can be generalized at three--loops and beyond without any additional effort. The 
reader is refered to \cite{Bierenbaum:2010cy} for further details.

\section{Duality Relation for multiple poles}
\label{sec:double-poles}

\begin{figure}[t]
\begin{center}

\begin{picture}(200,200)(0,-10)
\SetWidth{1.2}

\BCirc(100,100){70}
\Line(100,170)(100,30)

\ArrowArcn(100,100)(55,220,180)
\ArrowArc(100,100)(55,-40,0)

\ArrowLine(65,161)(50,187)
\ArrowLine(39,135)(13,150)
\ArrowLine(65,39)(50,13)
\Vertex(21,79){1.4}
\Vertex(29,59){1.4}
\Vertex(18,100){1.4}

 \Text(35,13)[]{$p_{N}$}
 \Text(36,187)[]{$p_{1}$}
 \Text(3,159)[]{$p_{2}$}

\ArrowArc(100,100)(70,0,30)

\ArrowArcn(100,100)(70,120,90)
\ArrowArcn(100,100)(70,150,120)
\ArrowArcn(100,100)(70,180,150)
\ArrowArcn(100,100)(70,270,240)

\ArrowLine(100,120)(100,80)

\Text(140,85)[]{$\ell_1$}
\Text(60,85)[]{$\ell_2$}
\Text(80,110)[]{$\ell_1+\ell_2$}

\end{picture}
\end{center}
\caption{\label{dpoleloop}
{\em The two--loop $N$--point scalar integral with a double pole}}
\end{figure}
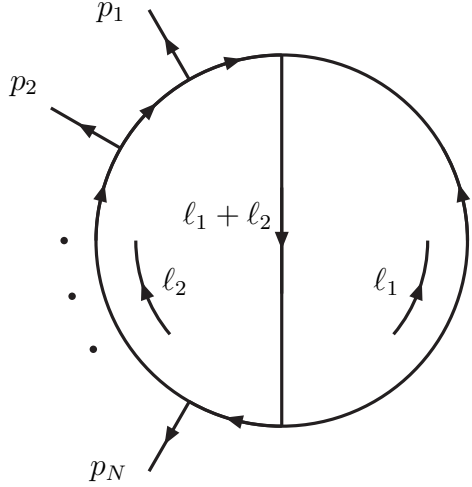

In the previous section we applied the Residue Theorem to one-- and two--loop graphs that
contain only single poles, i.e. no identical propagators. At one--loop this is always the case for a convenient
choice of gauge \cite{Catani:2008xa}. However,
at higher loops there exists the possibility of identical propagators, i.e. higher order
poles. Obviously, we need to generalize the Duality Theorem to accomodate for such graphs.
The first occurence of higher order poles is at the two--loop
level, with the sole double pole generic graph shown in Fig.~\ref{dpoleloop}. The Duality Theorem
can be derived for such graphs as well, using the Residue Theorem for multiple poles
\begin{align}
\text{Res}_{\{z=z_0\}}f(z)=\left.\frac{1}{(k-1)!}\left(\frac{d^{k-1}}{dz^{k-1}}(z-z_0)^{k}f(z)\right)\right|_{z=z_0}~.
\end{align}
The derivation follows similar steps as with the single pole case and is independent of any particular
coordinate system. We will derive an expression both in cartesian and light--cone coordinates, to demonstrate
this independence. We start with the cartesian system.
We write the Feynman propagator in a form that makes the poles explicit, i.e,
\begin{align}
G_F(q_i)=\frac{1}{(q_{i0}-q_{i0}^{(+)})(q_{i0}+q_{i0}^{(+)})}~,
\end{align}
where $q_{i0}^{(+)}=\sqrt{\mathbf{q}_i^2+m_i^2-i0}$ is the position of the pole. Then, applying the Residue Theorem 
by selecting poles with negative imaginary part, we have
\begin{align}
\text{Res}_{\{\textrm{Im} \,  q_{i0}<0\}}G_F^2(q_i)=-\frac{2}{(2q_{i0}^{(+)})^3}=
-\int dq_0\frac{1}{2(q_{i0}^{(+)})^2}\delta_+(q_i^2-m_i^2).
\end{align}
The imaginary component of the new denominator $1/2(q_{i0}^{(+)})^2$ is irrelevant, because it is always a 
positive quantity. We refer the reader to \cite{Catani:2008xa} where the calculation for the case of simple poles 
is explained in more detail. Then, we assume the following Lorentz-invariant prescription of the residue
\begin{align}
\text{Res}_{\{\text{Im} \, q_{i0}<0\}}G_F^2(q_i)=-\int dq_0\frac{\eta^2}{2(\eta q_{i})^2}\delta_+(q_i^2-m_i^2)~,
\label{ResCart}
\end{align}
where $\eta^{\mu}=(\eta_0,\mathbf{0})$ is a future-like vector, $\eta_0>0$, in cartesian coordinates.
Contrary to the one loop case, where numerators depending on the loop momentum do not modify the duality prescription, 
in the two loop and higher orders cases the derivative in the residue calculation introduced by the higher order poles 
act on every single term in the numerator and also on the remaining propagators. 
Let $N(\alpha_k)$ be a function of a set of momenta $q_l$, with $l \in \alpha_k$. 
Then the residue of a double pole is given by
\begin{align}
\text{Res}_{\{\text{Im} \, q_{i0}<0\}} \left\lbrace G_F^2(q_i)\left(\prod\limits_{j\neq i}G_F(q_j)\right)N(\alpha_k) 
\right\rbrace =
\left.\frac{\partial}{\partial q_0}\frac{1}{(q_{i0}+q_{i0}^{(+)})^2}
\left(\prod\limits_{j\neq i}G_F(q_j)\right)N(\alpha_k)\right|_{q_{i0}=q_{i0}^{(+)}}\nonumber\\
=\left(\prod\limits_{j\neq i}G_D(q_i;q_j)\right)\frac{1}{(2q_{i0}^{(+)})^2}
\left[-\frac{1}{q_{i0}^{(+)}}-\sum\limits_{j\neq i}(2q_{j0})G_D(q_i;q_j)+\frac{\partial}{\partial q_0}\right]N(\alpha_k)~,
\end{align}
which can be written as
\begin{align}
\label{eq:doublepolecart}
\text{Res}_{\{\text{Im} \, q_{i0}<0\}} 
\left\lbrace G_F^2(q_i)\left(\prod\limits_{j\neq i}G_F(q_j)\right)N(\alpha_k) \right\rbrace =
\int dq_0\delta_+(q_i^2-m_i^2)\left(\prod\limits_{j\neq i}G_D(q_i;q_j)\right)\nonumber\\
\times\left[-\frac{\eta^2}{2(\eta q_i)^2}-\sum\limits_{j\neq i}\frac{\eta q_j}{\eta q_i}G_D(q_i;q_j)
+\frac{1}{2\eta q_i}\frac{\partial}{ \partial \eta q_i}\right]N(\alpha_k)~.
\end{align}

In light--cone coordinates we choose our coordinates such that in the plus component complex plane the poles 
with negative imaginary part are located at:
\begin{align}
q_{i+}^{(+)}=\frac{q_{i\perp}^2+m_i^2-i0}{2q_{i-}},\qquad\text{with}\quad q_{i-}>0~.
\end{align}
In these light cone coordinates the Feynman propagator reads:
\begin{align}
G_F(q_i)=\frac{1}{2q_{i-}(q_{i+}-q_{i+}^{(+)})}~,
\end{align}
and thus
\begin{align}
\text{Res}_{\left\lbrace \text{Im\,}q_{i0}<0 \right\rbrace }\theta(q_{i-})G_F^2(q_i)=0~,
\end{align}
which, at first sight, seems to contradict equation \Eq{ResCart}. This contradiction can be resolved by taking into 
account the fact that in light cone coordinates, the dual vector $\eta$ is lightlike and therefore $\eta^2=0$. 
Hence equation \Eq{ResCart} remains valid. Now, we are ready to calculate the residue of a double pole in light cone 
coordinates:
\begin{align}
\label{eq:double-lightcone}
&\text{Res}_{\{\text{Im\,}q_{i0}<0 \}} 
\left\lbrace \theta(q_{i-})G_F^2(q_i)\left(\prod\limits_{j\neq i}G_F(q_j)\right)N(\alpha_k) \right\rbrace =
\left.\frac{\theta(q_{i-})}{(2q_{i-})^2}\frac{\partial}{\partial q_+}
\left(\prod\limits_{j\neq i}G_F(q_j)\right)N(\alpha_k)\right|_{q_{i+}=q_{i+}^{(+)}}\nonumber\\
&=\int dq_+\delta_+(q_i^2-m_i^2)\left(\prod\limits_{j\neq i}G_D(q_i;q_j)\right)
\left[-\sum\limits_{j\neq i}\frac{\eta q_j}{\eta q_i}G_D(q_i;q_j)+\frac{1}{2\eta q_i}
\frac{\partial}{\partial\eta q_i}\right]N(\alpha_k)~,
\end{align}
where now $\eta^{\mu}=(\eta_+,\eta_-=0,\mathbf{0}_{\perp})$. \Eq{eq:double-lightcone} has the same
functional form as in \Eq{eq:doublepolecart}, although with a different dual vector $\eta$. Thus we can
generalize \Eq{eq:doublepolecart} and \Eq{eq:double-lightcone} to an arbitrary coordinate system and
combining simple and double poles in a single formula we get in cartesian coordinates:
\begin{align}
\int_q G_F^2(q_i)&\left(\prod\limits_{j\neq i}G_F(q_j)\right)N(\alpha_k)=\nonumber\\
&-\int_q\Bigg\{\tilde{\delta}(q_i)\left(\prod\limits_{j\neq i}G_D(q_i;q_j)\right)
\left[-\frac{\eta^2}{2(\eta q_i)^2}-\sum\limits_{n\neq i}\frac{\eta q_j}{\eta q_i}G_D(q_i;q_j)
+\frac{1}{2\eta q_i}\frac{\partial}{\partial \eta q_i}\right]\nonumber\\
&+\sum\limits_{j\neq i}\tilde{\delta}(q_j)G_D^2(q_j;q_i)\left(\prod\limits_{k\neq i,j}G_D(q_j;q_k)\right)\Bigg\}
N(\alpha_k).
\label{doublepoleformula}
\end{align}

Equation (\ref{doublepoleformula}), is the main result of this section. It extends the Duality Theorem
to integrals with identical propagators or, to put it differently, with double poles in the complex plane.
For the case of the generic two--loop graph in Fig.~\ref{dpoleloop}, this result can be seen as an extension
of \Eq{eq:twoGD}. If we have two groups of momenta, $\alpha_k , \alpha_2$, one of which contains the 
double propagator, i.e. $\alpha_k=\left\lbrace q_n=\ell_1+\ell_2 \right\rbrace $ and 
$\alpha_2= \left\lbrace q_2=\ell_2,q_3=\ell_2+p_1,\ldots,q_{n-1}=\ell_2+p_{1,N-1},q_2=\ell_2  \right\rbrace $, 
and we denote by $\alpha_2^{\prime}$ a group that contains all the momenta of $\alpha_2$ leading to single poles, namely
\newline
$\alpha_2^{\prime}=\left\lbrace q_2=\ell_2,q_3=\ell_2+p_1,\ldots,q_{n-1}=\ell_2+p_{1,N-1} \right\rbrace$,
then we can write:
\bea
G_D(\alpha_k \cup \alpha_2)&=& \tilde{\delta}(q_2)\left(\prod_{j \in \alpha_2^{\prime},\alpha_k }^{n}G_D(q_2;q_j)\right)
\left[-\frac{\eta^2}{2(\eta q_2)^2}-\sum_{j \in \alpha_2^{\prime},\alpha_k}^{n}\frac{\eta q_j}{\eta q_2}G_D(q_2;q_j)
\right] \nn \\
&+&\sum_{i \in \alpha_2^{\prime},\alpha_k}^{n}\tilde{\delta}(q_i)G_D^2(q_i;q_2)
\left(\prod\limits_{j \neq i}^{n} G_D(q_i;q_j)\right).
\eea
This result states that for the case of a double pole, one follows the usual procedure of cutting every propagator
line once, including the double propagator, and transforming the rest of the propagators to dual propagators.
A similar formula can be derived for the case of multiple (triple and higher) poles. 
The calculation of the residue of a multiple pole introduces, however, contributions with powers of dual propagators.
In absence of a general transformation formula analogous
to \Eq{eq:GAinGDGeneralN}, it is not possible to rewrite \Eq{doublepoleformula} in terms
of dual propagators whose dual $+i0$ prescription depends
on the external momenta only.  For that reason, we will present in the next section a different strategy 
for dealing with higher order poles based on the reduction of the integral using Integration By Parts.

\section{Reducing to single poles with IBPs}
In this section, we discuss a different approach to the generalization of the Duality Theorem to higher order
poles. We will use Integration By Parts (IBP) \cite{Chetyrkin,Smirnov}
to reduce the integrals with multiple poles to ones with simple poles. We emphasize the fact 
the \emph{we do not need to reduce the integrals to a particular integral basis}. 
We just need to reduce them "enough", so that the higher order poles disappear.

To give a short introduction to the method and establish our notation, let us consider a general $m$--loop scalar integral
in $d$ dimensions, with $n$ denominators $D_1,\ldots,D_n$ raised to exponents $a_1,\ldots,a_n$ 
and external momenta $p_1,\ldots,p_N$:
\beq
\int_{\ell_1}  \cdots \int_{\ell_m}  \frac{1}{D_1^{a_1} \cdots D_n^{a_n}} \; .
\eeq
If we notice that
\beq
\int_{\ell_1}  \cdots \int_{\ell_m}   
\frac{\partial}{\partial s^\mu}\frac{t^\mu}{D_1^{a_1} \cdots D_n^{a_n}} = 0 \; ,
\label{eq:IBPdef}
\eeq
where $s^\mu=\ell_1^\mu, \ell_2^\mu, \ldots, \ell_m^\mu$, the integrand being a total derivative with
respect to the loop momenta, we can find relations between scalar integrals with different exponents
$a_i$. This will allow us to express integrals with exponents larger than one, in terms of simpler ones. In effect,
we will be able to write integrals with multiple poles in terms of sums of integrals with simple poles.
In the numerator of the integrand of \Eq{eq:IBPdef} we can use 
$t^\mu=\ell_1^\mu, \ldots, \ell_m^\mu,p_1^\mu,\ldots,p_{N}^\mu$, to obtain a
system of equations that relate the various integrals. For simplicity, when refering to an IBP we will
use the shorthand notation:
\beq
\frac{\partial}{\partial s} \cdot t
\eeq
to denote Eq.~(\ref{eq:IBPdef}).
The differentiation will raise or leave an exponent
unchanged, while, contractions with the loop and external momenta in the numerator of the integrand, 
can be expressed in terms
of the propagators to lower an exponent. Often times though, this is not possible, leaving scalar products
of momenta, which cannot be expressed in terms of denominators. These are called Irreducible Scalar Products (ISP).
We will consider ISPs as additional denominators, $D_{ij}=\ell_i \cdot p_j$ \, , with a negative index $a_i$. 
We use the notation:
\beq
F(a_1 a_2 \cdots a_n)= \int_{\ell_1} \int_{\ell_2} \frac{1}{D_1^{a_1}D_2^{a_2} \cdots D_n^{a_n}}
\eeq
to denote a generic two--loop integral with $n$ propagators raised to an arbitrary integer power, with 
$D_i=G_F^{-1}(q_i)=q_i^2-m_i^2+i0$ and $q_i$ denotes any combination of external and loop momenta. In the following the 
prescription $+i0$ for the propagators is understood. 
We will use the symbol ${\bf a_i^+}$ to denote the raising of the index $a_i$ by one i.e. 
${\bf 1^+}F(a_1, a_2, \cdots a_n)=F(a_1+1, a_2, \ldots ,a_n)$ and the symbol ${\bf a_i^-}$ 
to denote the lowering of the index $a_i$ by one i.e. 
${\bf 2^-}F(a_1, a_2, \cdots a_n)=F(a_1, a_2-1, \ldots ,a_n)$. A combination of the two means that the operators
apply at the same time i.e. ${\bf 1^+ 2^-}F(a_1, a_2, \cdots a_n)=F(a_1+1, a_2-1, \ldots ,a_n)$.
In the following we will use two automated codes, for the reduction, {\tt FIRE} \cite{smirnovfire}, 
a {\tt MATHEMATICA} package for the reduction of integrals and {\tt REDUZE 2} \cite{vonManteuffel:2012yz}
\footnote{Since the most obvious first approach seems to be to try to express the
  integrals with multiple poles in terms of the same integrals with
  only single poles, c.f. \Eq{eq:firstexample}, we used, in addition
  to the ``usual'' version of {\tt REDUZE 2}, in some cases a special
  patch for {\tt REDUZE 2} which provides a modification of its integral
  ordering in the final result. This modified version of {\tt REDUZE 2}
  delivered the results for the integrals in this desired form stated
  in the subsequent sections, while we used the normal version of the
  integral ordering for the remaining cases.Note
  that we also calculated explicitely the relations obtained from the modified version, in the easiest cases of the
  massless two-- and three--loop integrals which can be built by
  insertion of the the massless one--loop two--point function, and
  found agreement.} ,
a package written in C++, using {\tt GiNaC} \cite{ginac}.

\subsection{The case for two--loop diagrams}
The only generic two--loop scalar graph with $N$-legs and a double propagator is shown in Fig.~\ref{dpoleloop}. 
The simplest case is the two--point function with massless internal lines. 
The denominators are:
\beq
D_1=\ell_1^2 \; \; , \; \;  D_2=\ell_2^2 \; \; , \; \; D_3=(\ell_2+p)^2 \; \; , \; \; D_4=(\ell_1+\ell_2)^2 
\; \; , \; \; D_5 = \ell_1 \cdot p \, , \nn
\eeq
where we have introduced an ISP as an additional denominator 
\footnote{For {\tt REDUZE 2} the corresponding propagator is added and used as input instead.}.
In our notation, the integral we want to reduce is $F(12110)$ and to this end we use the six total derivatives 
\beq
\frac{\partial}{\partial \ell_i} \cdot \ell_j \; , \; \frac{\partial}{\partial \ell_i} \cdot p
\; \; , \;\; i,j=1,2.
\label{eq:IBP2loop2p}
\eeq
Applying these IBPs on $F(a_1a_2a_3a_4a_5)$ we get a system of recursive equations. Using specific 
values for the exponents $a_i$ we can solve this system and obtain $F(12110)$. For this particular case, 
we solve the system explicitely and the reader is referred
to the appendix \ref{app:systemsolve} for details. Finally we arrive at:
\beq
\label{eq:firstexample}
F(12110)=\frac{-1+3\epsilon}{(1+\epsilon)s} \; F(11110) \, ,
\eeq
where $s=p^2+i0$, a result which contains only single poles and can be treated using the Duality Theorem 
\cite{Bierenbaum:2010cy}. 
For the rest of the 
cases below and in the three--loop case in the next section, we have used {\tt FIRE} and {\tt REDUZE 2}
to perform the reductions and check our results.
For three external legs $p_1^2=p_2^2=0,p_3^2=(p_1+p_2)^2$ and massless internal lines, we have the denominators: 
\beq
D_1=\ell_1^2\; \; , \; \;D_2=\ell_2^2 \; \; , \; \; D_3=(\ell_2+p_1)^2 \; \; , \; \; D_4=(\ell_2+p_1+p_2)^2 \; \; , \; \;
D_5=(\ell_1+\ell_2)^2  \; \; , \; \; \nn
\eeq
\beq
D_6= \ell_1 \cdot p_1 \; \; , \; \; D_7=\ell_1 \cdot p_2  \; , \nn
\eeq
where, the last two are the ISPs that appear in this case. The integral we want to reduce is $F(1211100)$.
We use eight IBPs:
\beq
\frac{\partial}{\partial \ell_i} \cdot \ell_j \; , \; \frac{\partial}{\partial \ell_i} \cdot p_j
\; \; , \;\; i,j=1,2.
\label{eq:IBP2loop3p}
\eeq
A similar analysis to the one above, gives:
\beq
 F(1211100)=\frac{3\epsilon}{(1+\epsilon)s} F(1111100)=
-\frac{3(1-3\epsilon)(2-3\epsilon)}{\epsilon (1+\epsilon)s^3} F(1001100) \; ,
\eeq
where $s=(p_1+p_2)^2+i0$,
which, again contains only single poles and can be treated with the Duality Theorem. 

The inclusion of masses does not affect the general picture of the reduction. It solely
introduces numerators in some integrals after the reduction is done. But, as
we have stressed already, the application of the Duality Theorem is not affected by numerators since
it only operates on denominators \cite{Bierenbaum:2010cy}. As an illustrative example, let us consider the two--loop
graph with two external legs and one massive loop (see Fig.~\ref{dpoleloop}). For the case of the left
loop being massive (related to $\ell_2$), with mass $m$, the denominators involved are
\beq
D_1=\ell_1^2 \; \; , \; \; D_2=\ell_2^2-m^2 \; \; , \; \; D_3=(\ell_2+p)^2-m^2 \; \; , \; \; 
D_4= (\ell_1+\ell_2)^2-m^2 \; ,\nn
\eeq
with the addition of the irreducible scalar product
\beq
D_5= \ell_1 \cdot p \; ,
\eeq
needed to perform the reduction. Using the same IBPs of \Eq{eq:IBP2loop2p}, the result of the reduction, with
{\tt FIRE} is:
\bea
F(12110)&=&\frac{(\epsilon-1) \left[  -\epsilon s^2 +2 m^2 (9 \epsilon -2 \epsilon^2 -3) s +
   4 m^4 (-3+2\epsilon)(-1+2\epsilon)  \right]  }{2 \epsilon (2
   \epsilon-1) m^4 s \left(4 m^2-s\right)^2} F(00110) \nn \\
&+&\frac{A }{2 \epsilon (2 \epsilon-1) m^4 s \left(4m^2-s\right)^2}F(10110) 
-\frac{(\epsilon-1) }{2 (2 \epsilon-1) m^4 s} F(1-1110)  \nn \\
&-&\frac{(\epsilon-1)^2 \left(2 m^2-s\right)}{(2 \epsilon-1) m^4 s \left(4 m^2-s\right)} F(01010) 
-\frac{(\epsilon-1) \left(4 \epsilon m^2+2 m^2-s\right)}{2 (2 \epsilon-1) m^4 \left(4 m^2-s\right)} F(01110) \nn \\
&+&\frac{2 (\epsilon-1) \left(m^2-s\right) \left(10 \epsilon m^2-\epsilon s-3
   m^2\right) }{\epsilon (2 \epsilon-1) m^4 s \left(4 m^2-s\right)^2} F(1011-1) ,
\eea
with $s=p^2+i0$ and 
\beq
\label{eq:A}
A=(1-\epsilon) \left[ \epsilon s +2 (3-8\epsilon)m^2  \right]s^2 
+ 2 (1-2\epsilon) m^4 \left[2(3-4\epsilon)m^2-(6-5\epsilon)s \right] \; .
\eeq
The reduction generates two integrals with a numerator, namely
\bea
F(1-1110)&=&\int_{\ell_1}\int_{\ell_2} \frac{\ell_2^2-m^2}{D_1D_3D_4} \; , \nn \\
F(1011-1)&=&\int_{\ell_1}\int_{\ell_2} \frac{\ell_1 \cdot p}{D_1D_3D_4} \; , \nn
\eea
but the double poles have now disappeared. The result with {\tt REDUZE 2} reads:
\bea
  F(1 2 1 1 0)&=&
    -
    \frac{(\epsilon-1) \left(4 \epsilon m^2+2
   m^2-s\right)}{2 (2 \epsilon-1) m^4 \left(4
   m^2-s\right)}F(0 1 1 1 0) \nn \\
    &+&
     \frac{3 (\epsilon-1) \left(8 \epsilon m^4-12 \epsilon
   m^2 s+\epsilon s^2-4 m^4+4 m^2 s\right)}{2
   \epsilon (2 \epsilon-1) m^4 s \left(4 m^2-s\right)^2}F(1 -1 1 1 0) \nn \\
    &+&
     \frac{A}{2 \epsilon (2
   \epsilon-1) m^4 s \left(4 m^2-s\right)^2}F(1 0 1 1 0) \nn \\
    &+&
     \frac{(\epsilon-1) \left(8 \epsilon^2 m^2 s-2 \epsilon^2 s^2-16
   \epsilon m^4+6 \epsilon m^2 s+\epsilon s^2+12
   m^4-6 m^2 s\right)}{2 \epsilon (2 \epsilon-1)
   m^4 s \left(4 m^2-s\right)^2}F(0 1 0 1 0) \; , \nn \\
\eea
where $A$ is given by \Eq{eq:A}. Despite the appearence of
different integrals the two results are of course equivalent. This is because, the integrals $F(00110)$ and
$F(01010)$, in the result obtained with {\tt FIRE}, are identical (as can be seen by shifting the loop momenta), 
so the sum of their coefficients gives exactly the coefficient of the result obtained with  {\tt REDUZE 2}. 
The same argument applies for the integrals $F(1011-1)$ and $F(1-1110)$.
The appearence of the numerators does not affect the application
of the Duality Theorem for integrals with single poles as was detailed in \cite{Bierenbaum:2010cy}.
For the case of the right loop in Fig.~\ref{dpoleloop} being massive (related to $\ell_1$), we have the denominators:
\beq
D_1=\ell_1^2-m^2 \; \; , \; \; D_2=\ell_2^2 \; \; , \; \; D_3=(\ell_2+p)^2 \; \; , \; \; 
D_4= (\ell_1+\ell_2)^2-m^2 \; \; , \; \; D_5= \ell_1 \cdot p \; . \nn
\eeq
Using the IBPs from \Eq{eq:IBP2loop2p}, we get with {\tt FIRE}:
\bea
F(12110)&=&\frac{\left(32 \epsilon^2 m^4+8 \epsilon^2 m^2 s+\epsilon^2 s^2-32 \epsilon m^4-11 \epsilon m^2 s-\epsilon s^2+6 m^4+3 m^2 s\right) }{6 m^4
   s^2} F(10110) \nn \\ 
&-&\frac{(\epsilon-1) \left(16 \epsilon^3 m^2+4 \epsilon^3 s-20 \epsilon^2 m^2-8 \epsilon m^2-7 \epsilon s+3 m^2+3 s\right) }{6 (2 \epsilon-1)
   (2 \epsilon+1) m^4 s^2} F(10010) \nn \\
&-&\frac{(\epsilon-1) \epsilon }{3 m^4 s} F(1011-1) -\frac{(\epsilon-1) (2 \epsilon-1) }{2 m^2
   s} F(01110) \nn \\
&-&\frac{(\epsilon-1) \left(12 \epsilon m^2+\epsilon s-3 m^2\right) }{6 m^4 s^2} F(1-1110)\nn \\
&-&\frac{(\epsilon-1) \left(6 \epsilon m^2+\epsilon s-3
   m^2\right) }{6 m^4 s} F(11100) \; , 
\eea
and with {\tt REDUZE 2}:
\bea
  F(1 2 1 1 0)&=&
     -\frac{(\epsilon-1) \left(8 \epsilon m^2+\epsilon s-2
   m^2\right)}{4 m^4 s^2}
     F(1 -1 1 1 0) \nn \\
    &+&
    \frac{64 \epsilon^2 m^4+16 \epsilon^2 m^2 s+\epsilon^2
   s^2-64 \epsilon m^4-22 \epsilon m^2 s-\epsilon
   s^2+12 m^4+6 m^2 s}{12 m^4 s^2}F(1 0 1 1 0)\nn \\
    &-&\frac{(\epsilon-1) \left(12 \epsilon m^2+\epsilon s-6
   m^2\right)}{6 m^4 s}F(1 1 1 0 0)\nn \\
   &-&\frac{(\epsilon-1) (2 \epsilon-3) \left(16 \epsilon^2 m^2+2
   \epsilon^2 s+4 \epsilon m^2+3 \epsilon s-2 m^2-2
   s\right)}{12 (2 \epsilon-1) (2 \epsilon+1) m^4 s^2}F(1 0 0 1 0) \; .
\eea
For the case of the double pole, two--loop graph, with three external legs and one massive loop,
we have the denominators:
\beq
D_1=\ell_1^2 \; \; , \; \; D_2=\ell_2^2-m^2 \; \; , \; \; D_3=(\ell_2+p_1)^2-m^2 \; \; , \; \; 
D_4=(\ell_2+p_1+p_2)^2-m^2 \; \; , \; \; D_5=(\ell_1+\ell_2)^2-m^2 \; ,  \nn 
\eeq 
\beq
D_6= \ell_1 \cdot p_1 \; \; , \; \; D_7= \ell_1 \cdot p_2 \; . \nn
\eeq
Using the IBPs from \Eq{eq:IBP2loop3p}  we get with {\tt FIRE}:
\bea
F(1211100)&=&\frac{(\epsilon-1) 
\left\lbrace  (1+4\epsilon) s^2 -4 \epsilon m^2 s (11-2\epsilon) -8m^4 (4\epsilon^2-8\epsilon-1) \right\rbrace  }
{8 \epsilon (2 \epsilon-1) m^6 s \left(4 m^2-s\right)^2} F(0001100)\nn \\
&-&\frac{A_1}{2 \epsilon (2 \epsilon-1) m^6 s \left(4
   m^2-s\right)^2} F(1001100) \nn \\
&+& \frac{\left(8 \epsilon^3-12 \epsilon^2+4 \epsilon-1\right) (\epsilon-1) }
{8 \epsilon (2 \epsilon-1)^2 m^6 s} F(0010100)  \nn \\
&-&\frac{(\epsilon-1) }{2 (2 \epsilon-1) m^4} \left[ F(0111100) - F(1011100) \right]  
+\frac{2(\epsilon-1) }{m^2 s \left(4 m^2-s\right)} F(0101100) \nn \\
&+&\frac{(\epsilon-1) \left(2 \epsilon m^2-\epsilon s-m^2\right)
   }{2 \epsilon (2 \epsilon-1) m^6 s^2} F(1-101100) \nn \\
&+&\frac{(\epsilon-1)^2 \left(8 \epsilon m^2-2 \epsilon s-6 m^2+s\right)}{2 (2 \epsilon-1) m^6 s \left(4 m^2-s\right)} F(0100100) \nn \\
&-&\frac{2 (\epsilon-1) \left(m^2-s\right) 
\left\lbrace  -\epsilon s^2 +m^2 s (6\epsilon-1)+8m^4 (2\epsilon-1)  \right\rbrace  }
{\epsilon (2 \epsilon-1) m^6 s^2 \left(4 m^2-s\right)^2} F(10011-10) \; ,
\eea
where $s=(p_1+p_2)^2+i0$,  and:
\beq
A_1=\epsilon(1-\epsilon) s^3 + m^2 (-1+\epsilon)(-1+9\epsilon)s^2 +
  m^4 (1-\epsilon)(5+6\epsilon)s +2 m^6 (1+2\epsilon)(-3+4\epsilon) \; ,
\eeq
while, with {\tt REDUZE 2}, we get:
\bea
 F(1 2 1 1 1 0 0)
   &=&
    -\frac{\epsilon-1}{2 (2 \epsilon-1) m^4}
    F(0 1 1 1 1 0 0) \nn \\
    &+&\frac{\epsilon-1}{2 (2 \epsilon-1) m^4}
    F(1 0 1 1 1 0 0) \nn \\
    &+&\frac{2 (\epsilon-1)}{m^2 s \left(4 m^2-s\right)}
    F(0 1 0 1 1 0 0) \nn \\
    &-&\frac{3 (\epsilon-1) \left(4 \epsilon m^4-8 \epsilon m^2 s+\epsilon
   s^2+2 m^4+m^2 s\right)}{2 \epsilon (2 \epsilon-1) m^6 s \left(4
   m^2-s\right)^2}
    F(1 -1 0 1 1 0 0) \nn \\
     &-&\frac{A_2}{2 \epsilon (2 \epsilon-1)
   m^6 s \left(4 m^2-s\right)^2}
    F(1 0 0 1 1 0 0) \nn \\
     &+&\frac{A_3}{4 \epsilon (2 \epsilon-1)^2 m^6 s \left(4
   m^2-s\right)^2}
    F(0 1 0 0 1 0 0) \; ,
\eea
where 
\bea
A_2&=&16 \epsilon^2 m^6-6 \epsilon^2 m^4 s+9 \epsilon^2 m^2
   s^2-\epsilon^2 s^3 \nn \\
&-&4 \epsilon m^6+\epsilon m^4 s-10 \epsilon m^2
   s^2+\epsilon s^3-6 m^6+5 m^4 s+m^2 s^2 \; ,
\eea
 and
\bea
A_3&=&(\epsilon-1) \left(128 \epsilon^4 m^4-64 \epsilon^4 m^2 s+8
   \epsilon^4 s^2-256 \epsilon^3 m^4+112 \epsilon^3 m^2 s- 12\epsilon^3 s^2+192 \epsilon^2 m^4 \right. \nn \\
&-& \left. 92 \epsilon^2 m^2 s+8 \epsilon^2
   s^2-40 \epsilon m^4+26 \epsilon m^2 s-\epsilon s^2-12 m^4+4 m^2
   s-s^2\right) \; .
\eea

The cases with additional external legs can be treated in a similar manner. It can always be reduced to sums of 
integrals with single propagators at the expence of introducing numerators. Although no formal proof exists,
in all cases studied so far it has been possible to reduce to integrals where only single poles appear. The generality
of this result seems plausible \cite{Gluza:2010ws}.

Our strategy is now clear. For a two--loop calculation, first we reduce all double pole graphs using
IBPs or any other method. The remaining integrals all contain single poles and can be treated using
the Duality Theorem at two--loops. The appearence of vector or tensor integrals does not spoil this strategy
since the Duality Theorem for single poles, affects only the denominators of the integrands.

\subsection{The case for three--loop diagrams}

\begin{figure}[!htb]
\begin{center}
\includegraphics[scale=0.7]{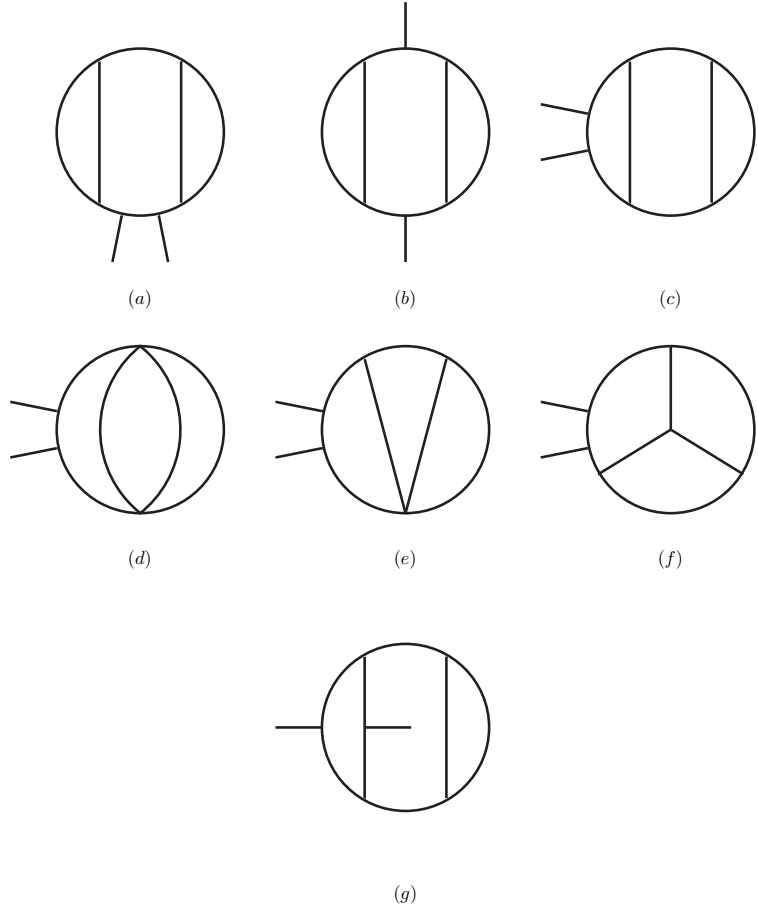}
\caption{\emph{Master topologies of three--loop scalar
    integrals with multiple powers of internal propagators.  
    Each internal line already dressed with some external 
    leg can be dressed with additional external lines.}}
\label{3LoopTop}
\end{center}
\end{figure}

For three--loop graphs there exists one topology with a triple propagator and a number of topologies
with a double propagator. All topologies are shown in Fig.~\ref{3LoopTop}. The arguments for the two--loop
case are valid here as well. We first reduce the multiple pole integrands by using IBPs until we have integrals with 
only single poles (possibly with numerators) and then we can then apply single-pole Duality Theorem as it was described
for the three--loop case in Ref. \cite{Bierenbaum:2010cy}. In the following,  we show explicitly the reduction 
of the two-point function for the different topologies and for massless internal lines.
We use the notation:
\beq
F(a_1 a_2 \cdots a_n)= \int_{\ell_1} \int_{\ell_2} \int_{\ell_3} \frac{1}{D_1^{a_1}D_2^{a_2} \cdots D_n^{a_n}}
\eeq
to denote a generic three--loop integral with $n$ propagators raised to an arbitrary integer power, with 
$D_i=G_F^{-1}(q_i)=q_i^2-m_i^2+i0$ or $D_i$ equal to any ISP and $q_i$  any combination of external and loop momenta. 
For the rest of this section the prescription $+i0$ for the propagators is understood. We also have $s=p^2+i0$.
The IBPs to be used for the reduction are:
\beq
\frac{\partial}{\partial \ell_i} \cdot \ell_j \; , \; \frac{\partial}{\partial \ell_i} \cdot p
\; \; , \;\; i,j=1,2,3
\label{eq:IBP3loop2p}
\eeq
In the following, we present first the result obtained with {\tt REDUZE 2} and then with {\tt FIRE}.
For the single triple pole graph (a) in Fig. (\ref{3LoopTop}), we have the following expressions:
\begin{description}
 \item[(a)] The denominators used are:
     \beq
       D_1=\ell_1^2 \; , \; D_2=\ell_2^2 \; , \; D_3=\ell_3^2 \; , \; D_4=(\ell_2-p)^2 \; , \; 
       D_5=(\ell_1-\ell_2)^2 \; , \; D_6=(\ell_3-\ell_2)^2 \; , \nn
     \eeq
     \beq
      D_7=\ell_1 \cdot p \; , \; D_8=\ell_3 \cdot p \; , \; D_9=\ell_1 \cdot \ell_3 \; , \nn
     \eeq
     with the result:
     \beq
      F(131111000)=\frac{2\epsilon(-1+4\epsilon)}{(1+\epsilon)(1+2\epsilon) s^2} F(111111000)=
       \frac{2(-1+2\epsilon)(-1+4\epsilon)}{ (1+\epsilon) (1+2\epsilon)s^3} \; F(101111000) \; .
     \eeq
 \end{description}
For the graphs with doubles poles, (b)-(g), Fig. (\ref{3LoopTop}), we find:
\begin{description}
 \item[(b)] The denominators are:
     \beq
       D_1=\ell_1^2 \; , \; D_2=\ell_2^2 \; , \; D_3=\ell_3^2 \; , \; D_4=(\ell_2-p)^2 \; , \; 
       D_5=(\ell_1-\ell_2)^2 \; , \; D_6=(\ell_3-\ell_2+p)^2 \; , \nn
     \eeq
     \beq
      D_7=\ell_1 \cdot p \; , \; D_8=\ell_1 \cdot \ell_3 \; , \; D_9=\ell_3 \cdot p \; , \nn
     \eeq
     with the result:
     \bea
      F(121211000)&=&
     \frac{3(-1+4\epsilon)(1+3\epsilon)}{(1+\epsilon)^2 s^2}F(111111000) \nn \\
      &=& \; 
      \frac{6(-2+3\epsilon)(-1+3\epsilon)(1+3\epsilon)(-3+4\epsilon)(-1+4\epsilon)}
       {\epsilon^2 (1+\epsilon)^2 (-1+2\epsilon) s^4}
      \; F(101011000)\; . \nn \\
     \eea
 \item[(c)] The denominators are:
     \beq
       D_1=\ell_1^2 \; , \; D_2=\ell_2^2 \; , \; D_3=\ell_3^2 \; , \; D_4=(\ell_3+p)^2 \; , \; 
       D_5=(\ell_3-\ell_2)^2 \; , \; D_6=(\ell_1-\ell_2)^2 \; , \nn
     \eeq
     \beq
      D_7=\ell_1 \cdot p \; , \; D_8=\ell_2 \cdot p \; , \; D_9=\ell_1 \cdot \ell_3 \; , \nn
     \eeq
     with the result:
     \bea
      F(122111000)&=&\frac{2\epsilon (-1+4\epsilon)(-1+3\epsilon)}
     {(1+2\epsilon)(1+\epsilon)^2 s^2}F(111111000) \nn \\
      &=& 
      \frac{2(-2+3\epsilon)(-1+3\epsilon)(-3+4\epsilon)(-1+4\epsilon)}{\epsilon (1+\epsilon)^2 (1+2\epsilon) s^4}
       \; F(100111000)\; .
     \eea
 \item[(d)] The denominators are:
     \beq
       D_1=\ell_1^2 \; , \; D_2=\ell_2^2 \; , \; D_3=\ell_3^2 \; , \; D_4=(\ell_3-p)^2 \; , \; 
       D_5= (\ell_2+\ell_3-\ell_1)^2  \; , \nn
     \eeq
     \beq
      D_6=\ell_1 \cdot p \; , \; D_7=\ell_2 \cdot p \; , \; D_8=\ell_1 \cdot \ell_2\; , \; D_9=\ell_1 \cdot \ell_3 \; , \nn
     \eeq
     with the result:
     \beq
      F(112110000)=\frac{(-1+2\epsilon)}{\epsilon s} \; F(111110000)
              =\frac{(-3+4 \epsilon)}{\epsilon s^2}  \; F(110110000)\; .
     \eeq
 \item[(e)] The denominators are:
     \beq
       D_1=\ell_1^2 \; , \; D_2=\ell_2^2 \; , \; D_3=\ell_3^2 \; , \; D_4=(\ell_3-p)^2 \; , \; 
       D_5=(\ell_1+\ell_2)^2 \; , \; D_6=(\ell_2+\ell_3)^2 \; , \nn
     \eeq
     \beq
      D_7=\ell_1 \cdot p \; , \; D_8=\ell_1 \cdot \ell_3 \; , \; D_9=\ell_2 \cdot p \; , \nn
     \eeq
     with the result:
     \beq
      F(112111000)=
       \frac{(-1+4\epsilon)}{(1+2\epsilon) s}F(111111000)
        = \frac{(-2+3\epsilon)(-3+4\epsilon)(-1+4\epsilon)}{\epsilon^2 (1+2\epsilon) s^3}
       \; F(100111000)\; .
     \eeq
 \item[(f)] The denominators are:
     \beq
       D_1=\ell_1^2 \; , \; D_2=\ell_2^2 \; , \; D_3=\ell_3^2 \; , \; D_4=(\ell_2+p)^2\; , \; 
       D_5=(\ell_1+\ell_2)^2 \; , \; D_6=(\ell_1+\ell_3)^2 \; , \nn
     \eeq
     \beq
     D_7=(\ell_3-\ell_2)^2\; , \;  D_8=\ell_1 \cdot p \; , \; D_9=\ell_3 \cdot p \; , \nn
     \eeq
     with the result:
     \bea
      F(121111100)&=&\frac{2\epsilon}{(1+\epsilon)s} F(111111100) \nn \\
      &=&
      \frac{2(-2+3\epsilon)(-1+3\epsilon)(-3+4\epsilon)(-1+4\epsilon)}{\epsilon^2 (1+\epsilon) (1+2\epsilon) s^4}
       \; \left[  F(001111000)+F(100101100) \right]  \nn \\
      &+& \frac{2(-1+2\epsilon)^2 (-1+4\epsilon)}{\epsilon (1+\epsilon) (1+2\epsilon)s^3} \; F(101110100)\; .
     \eea
 \item[(g)] The denominators are:
     \beq
       D_1=\ell_1^2 \; , \; D_2=\ell_2^2 \; , \; D_3=\ell_3^2 \; , \; D_4=(\ell_3-p)^2 \; , \nn
     \eeq
     \beq
       D_5=(\ell_1-\ell_2)^2 \; , \; D_6=(\ell_3-\ell_2)^2 \; , \; D_7=(\ell_3-\ell_2-p)^2 \nn
     \eeq
     \beq
      D_8=(\ell_1-\ell_3)^2\; , \;
      D_9=(\ell_1-p)^2
     \eeq
     with the result:
     \bea
     F(1 2 1 1 1 1 1 00 )
     &=&
    \frac{(-1+3 \epsilon)^2(1+5 \epsilon)}{(1+\epsilon)(1+2 \epsilon)^2 s^2}
     F(1 1 1 1 1 1 1 -1 0)\nn \\
    &+&\frac{\epsilon (9 \epsilon^2-11 \epsilon-4)}{(1+\epsilon) (1+2 \epsilon)^2  s^2}
     F(1 1 1 1 1 1 1 00)\; .
     \eea

\end{description}
The difference between the two results is due to the fact that the second is expressed in terms of
basis integrals while the first is expressed in terms of integrals with single poles of the 
same type as the multiple pole integral (in effect the
first result can be further reduced to the second). Since we do not seek a particular basis for our
reduction, as was stressed earlier, both results are equally useful as far as application of the Duality 
Theorem is concerned.

\section{Conclusions}
\label{sec:conclusions}

We have extended the Duality Theorem to two-- and
three--loop integrals with multiple poles. A Lorentz invariant expression
for the residues of double poles has been derived, which can
be extended straightforwardly to triple and, in general,
multiple poles. In the absence of a systematic procedure to
express the dual $+i0$ prescription in terms of external
momenta exclusively, as in the case of single poles,
we have explored an alternative approach.
We use IBP identities to reduce the integrals with identical propagators to ones with only single poles. 
Therefore, the essential features of the loop-tree duality now remain intact.  We reiterate that our goal 
is not to reduce everything to some set of master integrals. Rather, we reduce the integrals until 
there are no multiple poles left. Then, we can use the Duality Theorem in its original form for single pole
propagators, to rewrite them as integrals of a tree--level object over a modified
phase-space. The appearence of additional tensor integrals, due to the reduction, does
not affect our procedure, since applying the Duality Theorem in its single-pole version, only cuts propagators, leaving
the numerators of the integrals unaffected.

\section*{Acknowledgments} 

This work was supported in part by the Research Executive Agency
(REA) of the European Union under the Grant Agreement number PITN-GA-2010-264564
(LHCPhenoNet), 
by MICINN (FPA2011-23778, FPA2007-60323 and CSD2007-00042 CPAN) and by
GV (PROMETEO/2008/069), by the German Federal Ministry for Education
and Research BMBF through Grant No.\ 05~HT6GUA by the Helmholtz
Association HGF through Grant No.\ Ha~101 and by the German Research Foundation DFG through the 
Collaborative Research Center No 676 {\it Particles, Strings and the Early Universe-The Structure of Matter and 
Space-Time}. We would like to thank Stefano Catani for many elucidating discussions. I.B. would like to
thank Andreas von Manteuffel for providing us with the patch for {\tt REDUZE 2} that modified 
its  integral ordering for our purposes. 

\appendix

\section{Proof of the reduction of Eq. (\ref{eq:firstexample})}
\label{app:systemsolve}
Here we solve the system of equations, explicitely, to arrive at Eq. (\ref{eq:firstexample}). We note that we are
not aiming for a full reduction to a set of master integrals but rather to reduce the multiple poles to single
poles. Therefore, any integral which has single propagators is to be considered known. 

Using the IBPs, \Eq{eq:IBP2loop2p}, on the generic integral $F(a_1a_2a_3a_4a_5)$, we get the system of equations:
\beq
\label{eq:IBPsystem1}
d-2a_1-a_4-a_5 - a_4 {\bf 4^+} {\bf 1^-}+a_4 {\bf 4^+}{\bf 2^-}=0~,
\eeq
\beq
\label{eq:IBPsystem2}
a_1-a_4+ \frac{1}{2}s a_5 {\bf 5^+}+a_4 {\bf 4^+} ({\bf 1^-}-{\bf 2^-})+a_1 {\bf 1^+} ({\bf 2^-}-{\bf 4^-})
    +\frac{1}{2} a_5 {\bf 5^+} ({\bf 2^-}-{\bf 3^-})=0~, 
\eeq
\beq
\label{eq:IBPsystem3}
a_4 {\bf 4^+} (s+{\bf 2^-}-{\bf 3^-}-2 \; {\bf 5^-})-s a_5 {\bf 5^+}-2a_1 {\bf 1^+}{\bf 5^-}=0~, 
\eeq
\beq
\label{eq:IBPsystem4}
a_2-a_4+a_2 {\bf 2^+} ({\bf 1^-}-{\bf 4^-})+a_3 {\bf 3^+} ({\bf 1^-}+{\bf 2^-}-{\bf 4^-}-2 \, {\bf 5^-})+
   a_4 {\bf 4^+} ({\bf 2^-}-{\bf 1^-})=0~, 
\eeq
\beq
\label{eq:IBPsystem5}
d-2a_2-a_3-a_4 +a_3{\bf 3^+} (s-{\bf 2^-})+a_4 {\bf 4^+} ({\bf 1^-}-{\bf 2^-})=0~,
\eeq
\beq
\label{eq:IBPsystem6}
a_2-a_3+s  (a_2{\bf 2^+}-a_3{\bf 3^+}+a_4{\bf 4^+})+a_4 {\bf 4^+} ({\bf 2^-}-{\bf 3^-}-2 \, {\bf 5^-})
   +a_3{\bf 3^+ 2^-}-a_2{\bf 2^+ 3^-}=0~,
\eeq
where $s=p^2+i0$. The appearence of the operator ${\bf 5^-}$ signals that we have 
the ISP $\ell_1 \cdot p$ in the numerator of an integral. As long as these integrals possess single propagators,
we will not reduce them further but consider them known. We also note that a lot of the integrals that appear 
after setting particular values to the parameters $a_i$ in this system, are 
zero in dimensional regularization (in the massless case). 
Let us start by setting $a_2=2,a_1=a_3=a_4=1,a_5=-1$ in (\ref{eq:IBPsystem2}). We get:
\beq
\label{System1}
F(2111-1)-F(1112-1)-\frac{1}{2}sF(12110)-\frac{1}{2}F(11110)=0~.
\eeq
Taking the sum of (\ref{eq:IBPsystem5}) and (\ref{eq:IBPsystem6}) and setting $a_1=1,a_2=a_3=a_4=1,a_5=0$ we get:
\beq
\label{System2}
(d-4)F(11110)+sF(12110)+sF(11120)-2F(1112-1)=0~.
\eeq
Taking the difference between (\ref{eq:IBPsystem5}) and (\ref{eq:IBPsystem6}) 
and setting $a_1=2,a_2=a_3=a_4=1,a_5=0$ we get:
\beq
\label{System3}
-2F(2111-1)-sF(12110)+sF(11210)-F(10210)=0~.
\eeq
Finally, setting $a_1=2,a_2=a_3=a_4=1,a_5=-1$ in (\ref{eq:IBPsystem1}), we get:
\beq
\label{System4}
(d-4)F(2111-1)-F(1112-1)+F(2012-1)=0~.
\eeq
The integrals $F(10120)$ and $F(11210)$ and $F(2012-1)$, in this system of equations, can be computed simply by taking further, 
appropriate combinations of Eqs. (\ref{eq:IBPsystem1})-(\ref{eq:IBPsystem6}). Setting $a_1=a_2=a_3=a_4=1,a_5=0$ in
(\ref{eq:IBPsystem1}) we get $F(10120)=(3-d)F(11110)$. It also holds that $F(10120)=F(10210)$, as can be
seen by making the shifts in the loop momenta $\ell_2\rightarrow-\ell_1-\ell_2-p$.
From Eq. (\ref{eq:IBPsystem5}), by setting $a_1=a_2=a_3=a_4=1,a_5=0$ and using the value of  $F(10120)$, we get:
$s F(11210)=(10-3d)F(11110)$. Finally, adding Eqs. (\ref{eq:IBPsystem1}) and (\ref{eq:IBPsystem2}) and 
setting $a_1=a_2=1,a_3=2,a_4=1,a_5=-1$ we get: $F(2012-1)=(d-3)(d-4)F(11110)$. The rest of the system of equations
(\ref{System1})-(\ref{System4}) can now be solved sequencially, arriving at:
\beq
F(12110)= \frac{(3d-10)}{(d-6)s} F(11110) \, ,
\eeq
a result which, after putting $d=4-2\epsilon$, agrees with Eq. (\ref{eq:firstexample}).

\end{document}